\pdfoutput=1 
\documentclass[
  aps,%
  prl,%
 twocolumn,%
 groupedaddress,%
 superscriptaddress,%
  showpacs,%
 letterpaper,%
 amsfonts,%
 footinbib,%
 10pt,%
 floatfix,%
 noeprint%
]{revtex4-1}
\RequirePackage{amsmath,amssymb,bm}
\RequirePackage{graphicx}
\RequirePackage[usenames,dvipsnames]{xcolor}
\usepackage{pdfpages}
\usepackage{pgffor}
\RequirePackage{hyperref}
\hypersetup{%
  breaklinks = {true},
  citecolor = {blue},
  colorlinks = {true},
  linkcolor = {red},
  pdfauthor = {Su Pereira},
  pdfcreator = {\LaTeX\ and \flqq hyperref\frqq},
}

\graphicspath{ {./Figs/} }

\newcommand{\V}{{\cal{V}}}
\newcommand{\bk}{\bm{k}}

\newcommand{\bq}{\bm{q}}
\newcommand{\bQ}{\bm{Q}}
\newcommand{\meV}{\,\text{meV}}
\newcommand{\eV}{\,\text{eV}}
\newcommand{\Fref}[1]{Fig.~\ref{#1}}

\newcommand{\etal}{\emph{et al.}}
\newcommand{\KMO}{K$_{0.9}$Mo$_6$O$_{17}$}
\newcommand{\Tcdw}{\ensuremath{T_{\text{c}}}}
\newcommand{\Qcdw}{\ensuremath{\bm{Q}_{\text{cdw}}}}
\newcommand{\EgZero}{\ensuremath{E_g^0}}
\newcommand{\teq}{{\,=\,}}
\newcommand{\supmat}{supplementary information (SI)}

\renewcommand{\section}[1]{\emph{#1}\,---\,}

\makeatletter
\AtBeginDocument{\let\LS@rot\@undefined}
\makeatother

\begin{document}
\title{Charge Density Waves and the Hidden Nesting of Purple Bronze 
\texorpdfstring{\KMO}{K0.9Mo6O17}}

\author{Lei~Su}
\affiliation{%
  Centre for Advanced 2D Materials, National University of Singapore,
  6 Science Drive 2, Singapore 117546}

\author{Chuang-Han Hsu}
\affiliation{%
  Department of Physics, National University of Singapore,
  2 Science Drive 3, Singapore 117542}
\affiliation{%
  Centre for Advanced 2D Materials, National University of Singapore,
  6 Science Drive 2, Singapore 117546}
  
\author{Hsin Lin}
\affiliation{%
  Department of Physics, National University of Singapore,
  2 Science Drive 3, Singapore 117542}
\affiliation{%
  Centre for Advanced 2D Materials, National University of Singapore,
  6 Science Drive 2, Singapore 117546}
  
\author{Vitor~M.~Pereira}
\email{Corresponding author (vpereira@nus.edu.sg).}
\affiliation{%
  Department of Physics, National University of Singapore,
  2 Science Drive 3, Singapore 117542}
\affiliation{%
  Centre for Advanced 2D Materials, National University of Singapore,
  6 Science Drive 2, Singapore 117546}

\date{\today} 

\begin{abstract}
We introduce the first multiorbital effective tight-binding model to describe 
the effect of electron-electron interactions in this system. Upon fixing all the 
effective hopping parameters in the normal state against an \emph{ab initio} 
band structure, and with only the overall scale of the interactions as the sole 
adjustable parameter, we find that a self-consistent Hartree-Fock solution 
reproduces extremely well the experimental behavior of the charge density wave 
(CDW) order parameter in the full range $0{\,<\,}T{\,<\,}\Tcdw$, as well as the 
precise reciprocal space locations of the partial gap opening and Fermi arc 
development.
The interaction strengths extracted from fitting to the experimental CDW gap 
are consistent with those derived from an independent Stoner-type analysis. 
\end{abstract}

\maketitle
%
%

The layered purple bronze K$_{0.9}$Mo$_6$O$_{17}$ (KMO) has a triple-$\bQ$ charge density wave (CDW) 
phase below $\Tcdw\,{\simeq}\,120\,\text{K}$ \cite{greenblatt1988molybdenum} and
became the hallmark of ``hidden nesting'' \cite{whangbo1987band, 
whangbo1991hidden} because (i) despite its 3D layered structure, it has a 
strongly anisotropic Fermi surface (FS) topology, (ii) has a robust CDW phase below $\Tcdw$, 
(iii) does not develop a lattice distortion despite the commensurate CDW 
wave vector ($\Qcdw$) \cite{mou2016discovery}, and (iv) none of  the 
\emph{formal} Fermi sheets are individually nested by $\Qcdw$. 
Although the most recent experiments favor a purely electronically driven CDW 
instability \cite{mou2016discovery}, the K and Na purple bronzes remain largely 
unexplored theoretically. Despite seminal work by Whangbo \etal\ establishing 
the essential of the noninteracting electronic structure 
\cite{whangbo1991hidden}, there is no encompassing microscopic model 
that addresses the role of interactions and is capable of reproducing the key 
experimental observations associated with the CDW phase. This contrasts 
with the related Li$_{0.9}$Mo$_6$O$_{17}$, known to be quasi-1D and for which  
microscopic frameworks based on the Hubbard model \cite{merino2012effective} and 
Luttinger liquid theory \cite{chudzinski2012luttinger} have been proposed.

In a recent experiment, Mou \etal\ reported an outstanding difference between 
electronic states in the bulk and at the surface of KMO, inferred from 
observations by angle-resolved photoemission spectroscopy (ARPES) of a 
much higher $\Tcdw$ at the surface (${\simeq\,}220$\,K) \cite{mou2016discovery} 
and a tenfold increase of the associated ``surface CDW gap.'' These remarkable 
findings stress the urgency for a theoretical understanding of the mechanisms 
underlying such a large tunability of both $\Tcdw$ and $\EgZero$ in the same 
compound and, in particular, clarifying whether or not that arises from 
variations in the strength of the relevant interactions, and whether 
interactions are strong or weak.

\begin{figure}[tb]
\centering
\includegraphics[width=\columnwidth]{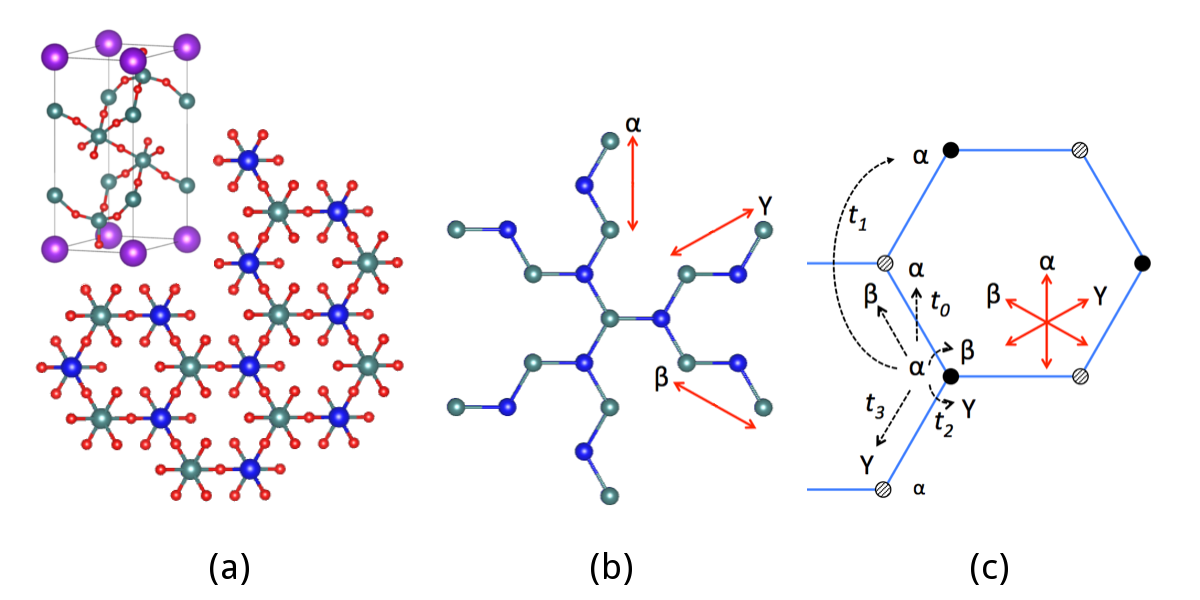} 
\caption{(a) Top view of the two inner KMO sublayers of composition Mo$_2$O$ _9$, and 
the full unit cell (inset); Mo, green and blue, O, red, K, purple. 
(b) Simplified representation of the slab shown in (a) with the oxygens 
removed and highlighting the \emph{effective} 1D zigzag chains generated by the 
three degenerate Mo $d$ orbitals.
(c) Diagram of the site, orbital and hopping labels used in our tight-binding 
model.
}
\label{fig:structure}
\end{figure}

We establish here an effective model for KMO that (i) accurately accounts for 
the noninteracting physics of the relevant $t_{2g}$-derived bands, (ii) 
identifies the dominant Coulomb interactions and their magnitudes, (iii) 
captures extremely well the temperature dependence of the CDW gap, and (iv) 
accurately reproduces the band folding, renormalized FS,  and Fermi arcs below 
$\Tcdw$.

\section{Noninteracting tight binding.}%
\begin{figure}[tb]
\centering
\includegraphics[width=\columnwidth]{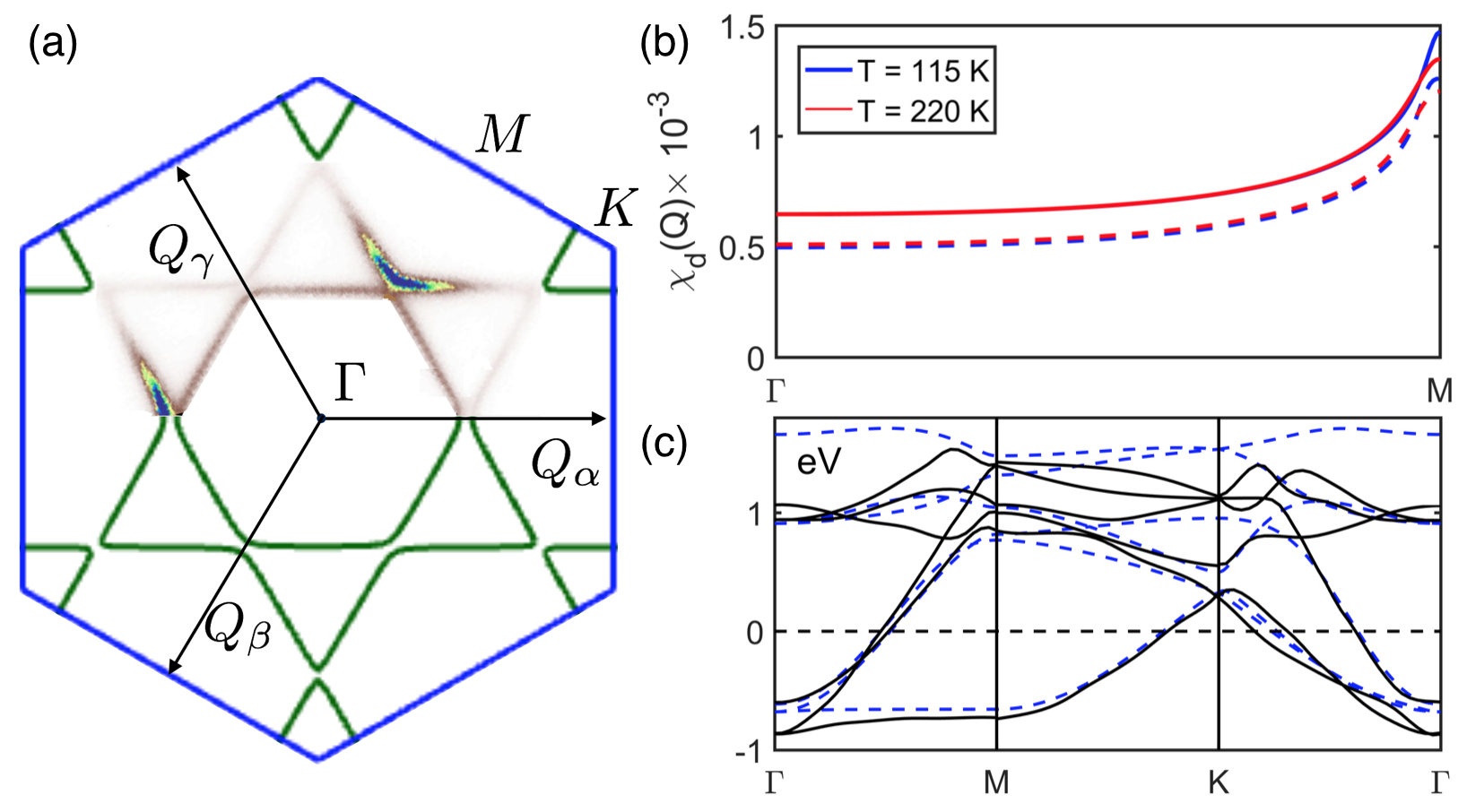} 
\caption{(a) Calculated Fermi contours in the normal state ($T{\,>\,}\Tcdw$, green 
lines) plotted together with the FS measured in Ref.~\cite{mou2016discovery}.
$\bQ_{\mu}$ represent the three experimentally measured CDW wave vectors. 
(b) Intrasublattice and intraorbital susceptibility ($\chi_d$) along 
$\Gamma M$ at $T\teq 115$ and $220$\,K. The dashed lines show $\chi_d$ 
calculated for decoupled 1D zigzag chains, while solid ones 
(vertically shifted by 0.1 for clarity) correspond to the full 2D TB model 
(see Supplemental Material for details \cite{supinfo}).
(c) Overlaid \textit{ab initio} (solid lines) and TB (dashed lines) bands.
}
\label{fig:FS}
\end{figure}

The crystal structure of KMO is illustrated in \Fref{fig:structure}(a) 
\cite{buder1982, vincent1983, greenblatt1988molybdenum}. The system is a good 
and strongly two-dimensional metal 
\cite{buder1982,degiorgi1988,xu2012transport}, understood as the result of the
oxygen-mediated overlap between $t_{2g}$-derived 
orbitals in the two inner slabs that leads to three half-filled bands 
\cite{buder1982,whangbo1987band}. The directional character of these effective 
overlaps leads to weakly hybridized quasi-1D Fermi sheets \cite{whangbo1987band} 
[\Fref{fig:FS}(a)] that underlie the \emph{hidden nesting} and its CDW 
instability \cite{whangbo1991hidden, gweon1997direct,valbuena2006charge}. 
Partial gaps are believed to develop below $\Tcdw $, since the system loses only
$50\%$ of its conductivity in the normal phase \cite{buder1982, degiorgi1988, 
xu2012transport, breuer1996observation}.

Figure~\ref{fig:structure}(a) shows that the network of Mo atoms in the electronically 
relevant two inner layers defines a honeycomb lattice. 
To make full use of the $C_3$ symmetry we introduce three \emph{effective} and 
equivalent orbitals: $\alpha,\,\beta,\,$ and $\gamma$. Their overlap reflects the 
effective $p$-mediated overlap between $t_{2g}$ orbitals with a $\pi$ ($\delta$) 
character along the intra-(inter) chain directions. Each orbital is associated 
with one of the three equivalent directions denoted by the red arrows in 
\Fref{fig:structure}(b). For example, in \Fref{fig:structure}(c), there is a 
sizable hopping between $\alpha$ orbitals along the upward-running zigzag 
chain, but a negligible one among $\alpha$ orbitals along the other two zigzag 
directions, and equivalently for $\beta$ and $\gamma$.
We consider the four hopping parameters represented in \Fref{fig:structure}(c): 
$t_0$ and $t_1$ account for the intraorbital hopping ($\alpha{-}\alpha$, 
$\beta{-}\beta$, and $\gamma{-}\gamma$) among nearest and next-nearest neighbors 
along the corresponding zigzag direction, respectively;
$t_2$ is an on-site interorbital hopping ($\alpha{-}\beta$, etc.); 
$t_3$ is a nearest-neighbor interorbital hopping on the bonds shared by the 
two corresponding chains (e.g., it represents the hopping between 
$\alpha{-}\beta$ and $\alpha{-}\gamma$).
The Fourier transform of this six-orbital tight-binding (TB) Hamiltonian is represented by
\begin{equation}
  H_0  = \sum_{\bm{k}\mu\nu I J} T_{\mu I, \nu J}(\bk)  \, 
    c^{\dagger}_{\mu I\bm{k}} \, c_{\nu J\bm{k}} 
  ,
  \label{H}
\end{equation}
where $c_{\mu I\bk}$ destroys an electron with crystal momentum $\bk$ at 
orbital $\mu{\,\in\,}\{\alpha,\beta,\gamma\}$ and sublattice 
$I{\,\in\,}\{A,B\}$. The explicit six-dimensional matrix $T_{\mu I, \nu J}(\bk)$ 
is provided in \supmat\ \cite{supinfo}, together with the details of the 
\emph{ab initio} calculations and an extended discussion of alternative 
approaches to obtain an appropriate TB model, such as through density-functional theory (DFT)-derived 
Wannier functions \cite{Marzari2012}. The TB parameters and chemical potential 
are determined by fitting the three partially filled bands to the DFT 
band structure within 1\eV\ of the Fermi level and ensuring the filling factor 
is preserved in the resulting TB \cite{supinfo}. We obtained 
$\{t_0,\,t_1,\,t_2,\,t_3,\,\mu\} \teq \{454,\,-204,\,136,\,114,\,659\}$\,meV.

Figure~\ref{fig:FS}(a) shows that this effective Hamiltonian precisely captures the 
experimental FS \cite{gweon1997direct,mou2016discovery} and 
reproduces the overall features of the three occupied DFT bands 
\Fref{fig:FS}(c).
In particular, the two inner electron and outer hole pockets arise from the 
avoided crossings of the underlying 1D FSs as a result of the small 
hybridization (controlled by $t_{2,3}$) between these effective 1D chains.
A na\"ive consideration of the Peierls instability for each independent 1D 
chain would not uniquely predict the CDW wave vectors in this system. 
Conversely, none of the individual 2D Fermi sheets is nested by the 
experimental $\bQ_\mu$, and the consideration of the nesting condition of each Fermi 
sheet independently would predict a different set of CDW wave vectors.
This ``contradiction'' between the \emph{apparent} nesting vectors of a 
strongly anisotropic 2D FS and the \emph{actual} $\bQ_{\mu}$ that 
describe the CDW is the essence of the hidden nesting concept 
\cite{whangbo1991hidden}: The experimentally observed $\bQ_\mu$ are preferred 
because each can simultaneously nest two of the three ``hidden'' 1D bands over 
the entire BZ.

\section{Coulomb interactions.}%
The good band structure fitting captured by the noninteracting model in 
\Fref{fig:FS}(a) suggests that interactions between quasiparticles are 
relatively small, at least in the normal state. These are introduced in the 
framework of a multiorbital extended Hubbard model, similar to the description 
of iron-based superconductors \cite{graser2009near}, where only the direct 
coupling terms are considered (no exchange):
\begin{equation}
  V = \frac{1}{2\V }\sum_{\bk} \sum_{\mu\nu I J}   V_{\mu I\nu J}(\bk) 
    \rho_{\mu I} (\bk ) \rho_{\nu J} (-\bk )
  .\label{V}
\end{equation}
$\V$ denotes the volume of the system, and $\rho_{\mu I}(\bq) = \sum_{\bk} 
c^{\dagger}_{\mu I \bk +\bq} c_{\mu I\bk } $ is the density operator. Both 
on-site and neighboring interactions include intraorbital Hubbard terms $U$ 
between electrons with different spins and interorbital Coulomb-like terms. We 
distinguish interactions along the chain directions (intrachain) and across 
adjacent chains (interchain) due to the anisotropy in the orbitals involved.
Our choice of three effective parameters captures the essential details of 
the electronic interactions in this system \cite{supinfo}: $V_1$ ($V_2$) defines 
intrachain intra(inter)-sublattice interactions (quasi-1D, along each 
equivalent zigzag), and the anisotropy factor $\eta$ determines the extent to 
which the full interacting Hamiltonian is more of a 1D nature ($\eta\teq 0$ for 
interactions only among orbitals belonging to the same chain) or more 
2D ($\eta > 0$).

\section{Stoner analysis and Hartree-Fock treatment.}%
To assess the magnitude of the interactions capable of driving the system into 
the CDW phase, we studied the generalized Stoner criterion for this 
instability along the same lines used, for example, in multiorbital iron-based 
superconductors \cite{graser2009near}. 
The RPA is used to obtain the strength of the interaction parameters 
compatible with the experimental $\Tcdw$ at $\bq\teq\Qcdw$ \cite{supinfo}.
\Fref{fig:FS}(b) shows the dominant diagonal element $\chi_d$ 
(\textit{intrasublattice} and \textit{intraorbital}) of the electronic 
susceptibility matrix as a function of the temperature and momentum along in two 
cases: the 1D limit of decoupled chains and the full 2D TB model (see Supplemental Material for 
details \cite{supinfo}). From the Stoner criterion, we estimate 
$280{\,\lesssim\,}V_1{-}V_2{\,\lesssim\,} 800\,\meV$. The upper bound is 
obtained in the limit of 1D-only interactions, $\eta\teq 0$, and the lower for 
isotropic interactions, $\eta\teq 1$. Since these are considerably smaller than 
the bandwidth, we may treat this as a weak or intermediate coupling system, 
justifying \emph{a posteriori} the analysis based on the RPA.
Note that the bare susceptibility curves plotted in \Fref{fig:FS}(b) at 
115 and 220\,K differ only slightly, which suggests that a small change 
in the interaction parameters (a different screening environment) can easily 
raise \Tcdw\ from 115 to 220\,K. 

In order to describe the temperature dependence of the CDW order parameter, we 
perform a Hartree-Fock mean field (MF) decoupling of the interactions in 
Eq.~(\ref{V}) and minimize the electronic free energy $F$
with respect to the order parameters $\Delta_{\mu I}(\bQ) {\,\equiv\,}  \langle \rho_{\mu I} (\bQ) \rangle $, where 
$\bQ {\,\in\,} \{\pm \bQ_{\alpha}, \pm \bQ_{\beta}, \pm \bQ_{\gamma} \}$ 
\cite{supinfo}. The minimization is done numerically due to the large 
24-dimensional structure of the decoupled Hamiltonian ($6{\times}4$).
 
\begin{figure}[t]
\centering
\includegraphics[width=\columnwidth]{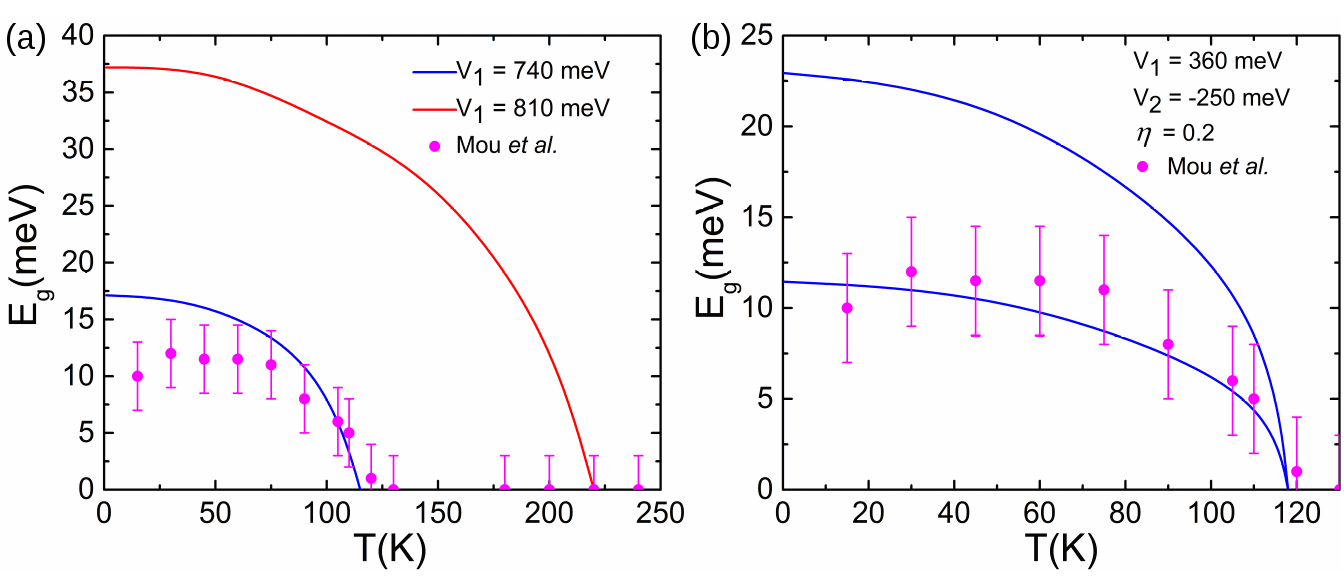} 
\caption{Gap along $\Gamma K'$ as a function of the temperature. (a) Simplified treatment 
of the interactions (intraorbital interactions, but complete 2D hoppings) 
discussed in the text. $V_1 \teq 740\,\meV$ and $V_1 \teq 810\,\meV$ yield
$\Tcdw\teq 115$\,K and $\Tcdw\teq 220$\,K, respectively. (b) The two gaps 
along $\Gamma K'$ for the more realistic interacting Hamiltonian [see 
also \Fref{fig:BS-SpeFun-DOS}(a)]. The experimental data (points) are from 
Mou \etal\,\cite{mou2016discovery}.
}
\label{fig:Gaps}
\end{figure}

Consider first the simpler case of interactions restricted to each chain 
(intraorbital interactions, but complete 2D hoppings). In this case, the MF 
solution depends only on the difference $V_1 -V_2$; we set $V_2 = 0$ and vary 
$V_1 >0$ until $\Tcdw$ is either 115 or 220\,K, in order to compare the 
results with the experimental transitions attributed to the  bulk and 
surface \cite{mou2016discovery}. 
The temperature dependence of the gap ($E_g$) along $\Gamma K'$ is presented 
in \Fref{fig:Gaps}(a).  
Since in this case the Coulomb interactions are determined by one 
effective parameter only, it is not surprising to find a BCS-like 
behavior in $E_g(T)$. Significantly, in order to make $\Tcdw \teq 
115$\,K, we must have $V_1{\,\approx\,} 740\,\meV$, in agreement with our 
independent estimate based on the generalized Stoner criterion. 
Moreover, the zero-temperature gap $\EgZero{\,\approx\,}17\,\meV$, consistent 
with the experimental value attributed to the bulk \cite{mou2016discovery}. In 
other words, three nearly independent zigzag chains seemingly suffice to explain 
well, quantitative and qualitatively, the bulk properties of KMO at the MF 
level. 
Figure~\ref{fig:Gaps}(a) also illustrates the high sensitivity of $\Tcdw$ to the 
magnitude of $V_1$, since a 10\% increase in the latter causes a twofold 
amplification of $\Tcdw$. Even though this suggests that $\Tcdw$ can be 
very sensitive to the local details of the interactions (screening, in 
particular) and might be easily placed at the values $\Tcdw{\,\approx\,} 220$\,K 
attributed to the surface, the associated low-temperature gap is far from the 
reported value of $150\,\meV$ \cite{mou2016discovery}. Conversely, setting 
$\EgZero \teq 150\,\meV$ ($V_1{\,\sim\,} 1050 \meV$) leads to 
$\Tcdw{\,\approx\,} 950$\,K. 

The more general, yet manageable, model of the Coulomb interactions introduces 
the three independent parameters $V_1$, $V_2$, and $\eta$ described earlier. 
Compared with the 1D limit, the CDW phase is now more stable as, for the same 
value of $\EgZero$, we obtain a larger \Tcdw. If $V_1 \teq 360\,\meV$, $V_2 \teq 
-250\,\meV$, and $\eta \teq 0.2$, we obtain perfect agreement with the 
experimentally reported values \cite{mou2016discovery}. The full temperature 
dependence shown in \Fref{fig:Gaps}(b) matches very well with the experimental 
data.
Note that this parameter set is still far from the isotropic limit and 
fulfills the Stoner bounds $280 {\,\lesssim\,} V_1 - V_2 {\,\lesssim\,} 800 
\meV$ obtained above. 

\begin{figure}[t]
\centering
\vspace*{-0.09in}
\includegraphics[width=\columnwidth]{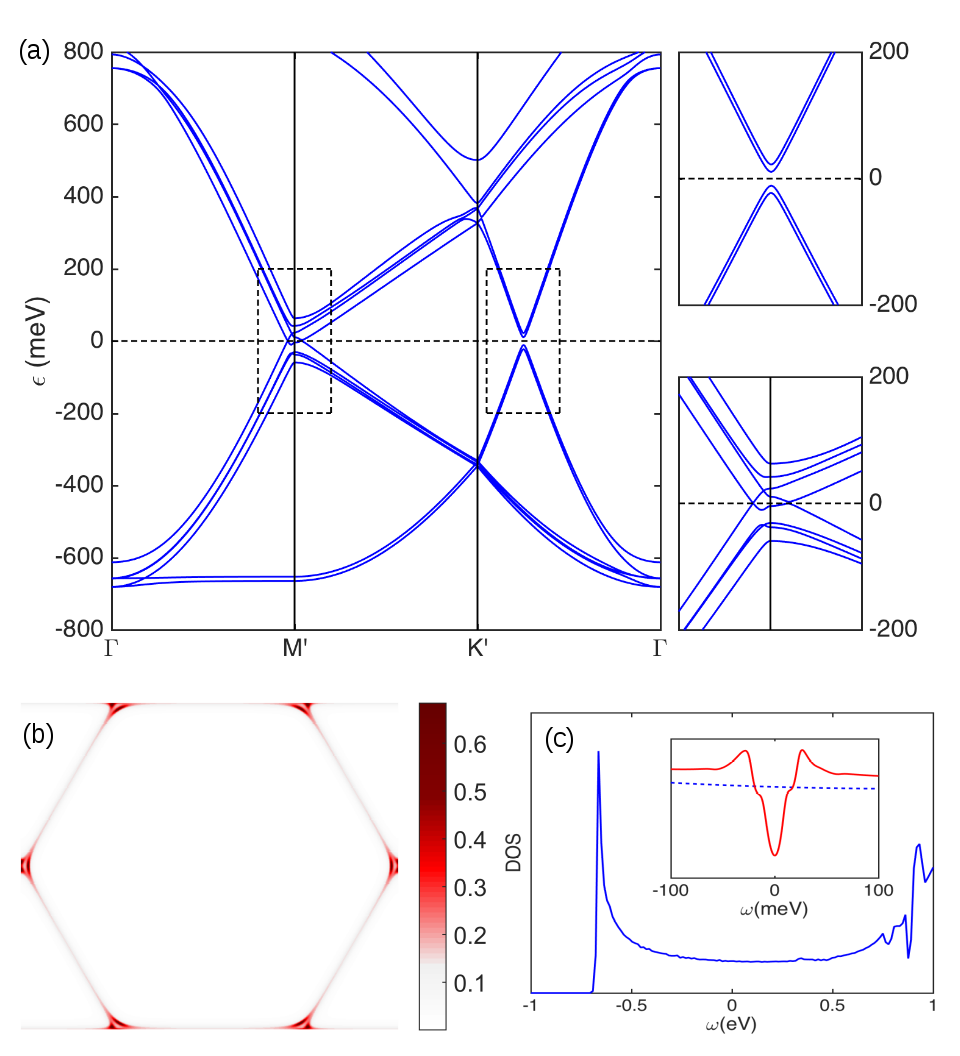} 
\caption{(a) Self-consistent band structure in the CDW phase along the high symmetry 
directions of the \emph{folded} zone at $T\teq 0$ ($V_1 \teq 360\,\meV$, 
$V_2 \teq -250\,\meV$, $\eta \teq 0.2$).
The rightmost panels amplify the dashed rectangles around $M'$ and $\Gamma K'$, respectively.
(b) Spectral function at $\omega \teq \mu$ (artificially broadened by 
$5\,\meV$, folded zone). Ungapped portions of the FS (Fermi arcs) 
lie around $M'$ (red, darker regions).
(c) Noninteracting (main panel) and renormalized (inset) DOS.
}
\label{fig:BS-SpeFun-DOS}
\end{figure}

Furthermore, the model captures the crucial fact that the gap opens at 
the right position along $\Gamma K'$ in the folded BZ. This is shown in 
detail in \Fref{fig:BS-SpeFun-DOS}(a), where, moreover, it is clear that the 
point $M'$ remains gapless despite a finite amount of repulsion among some of 
the folded bands. 
This ensures that the system undergoes a metal-metal transition, rather than 
metal-insulator, upon entering the CDW phase, in accord with transport 
experiments \cite{escribe1984evidence,xu2012transport}. The remaining 
electron-hole pockets at $M'$ in the CDW phase are also consistent with the 
experimental fact that charge carriers change from electron- to hole-like when 
entering the CDW \cite{buder1982,greenblatt1988molybdenum}.
To better illustrate the FS restructuring below $\Tcdw$, the 
spectral function at $\mu\teq 0$ is shown in \Fref{fig:BS-SpeFun-DOS}(b): The 
gaps along $\Gamma K'$ and the finite band overlaps at $M'$ create Fermi arcs 
centered at $M'$, compatible with the ARPES observations of enhanced spectral 
weight at these points \cite{gweon1997direct,mou2016discovery}.
The corresponding density of states (DOS) of the noninteracting model in the normal state is 
compared with the $\bk$-integrated spectral function in 
\Fref{fig:BS-SpeFun-DOS}(c). Whereas the former is nearly constant near $E_F$, 
the spectral function has a marked dip below $\Tcdw$, in qualitative agreement 
with earlier STM measurements \cite{mallet1999contrast} but remains finite as 
anticipated from the persistence of the Fermi arcs in 
\Fref{fig:BS-SpeFun-DOS}(b).

We note that this Fermi arc scenario is the one intuitively expected within the 
hidden nesting picture: Since the hybridization between the underlying 1D 
chains is strongest at $M'$ where they would otherwise be degenerate, the 
vicinity of this point is where the least nesting occurs among the 1D Fermi 
surfaces. The tendency for gap opening is strongest at $\Gamma K'$ (stronger 
nesting) than at $M'$. Moreover, the experimental FS at 
$T{\,\gtrsim\,}\Tcdw$ \cite{gweon1997direct,mou2016discovery} has a clear 
predominance of spectral weight at $M'$, consistent with 
\Fref{fig:BS-SpeFun-DOS}(b)
\footnote{
Nevertheless, we have explicitly verified that the model easily accommodates 
a fully gapped scenario where both $M'$ and $\Gamma K'$ become gapped below 
$\Tcdw$ by increasing $\eta$, which renders the interactions more 
two-dimensional.
}.

Figure~\ref{fig:BS-SpeFun-DOS}(a) also shows that there are, in fact, two ``gaps'' 
along $\Gamma K'$ because there are two quasidegenerate bands there in the 
normal state as a result of the BZ folding. Whereas in the 1D treatment of the 
interactions, the gap opens without lifting this degeneracy, the more 2D 
interaction lifts it and the two bands are pushed down by different amounts, as 
shown explicitly in \Fref{fig:BS-SpeFun-DOS}(a) and \ref{fig:BS-SpeFun-DOS}(b) and in \Fref{fig:Gaps}(b) as 
a function of the temperature. 
The ARPES data also reveal two gaps at this point, one attributed to the 
bulk and another that backbends below $\Tcdw$ attributed to the surface 
layers. It is tempting to relate them to the features of 
\Fref{fig:BS-SpeFun-DOS}(a) along $\Gamma K'$. However, 
on the one hand, the splitting of the two bands below $\Tcdw$ seen in 
\Fref{fig:BS-SpeFun-DOS}(a) cannot correspond to the two bands in the experiment, 
because the splitting of either of them follows a BCS trend as a function of the 
temperature [\Fref{fig:Gaps}(b)], unlike the surface gap that seems to set in 
instantaneously below $\Tcdw$. On the other hand, the second band in 
\Fref{fig:BS-SpeFun-DOS}(a) is a consequence of the BZ folding, and the 
experiments, despite the robust CDW, show no sign of backfolding in the bands 
attributed to the bulk of the system.
Hence, either the second band that is being pushed down along $\Gamma K'$ in 
\Fref{fig:BS-SpeFun-DOS}(a) lies further away from the Fermi level in the real 
system than with the parameters chosen in \Fref{fig:BS-SpeFun-DOS}(a), or the 
backfolded spectral weight is too weak to be detected experimentally, in which 
case this secondary gap would be discernible only in the extended zone, on the 
FSs cut by the BZ boundary, for example. Measurements along larger 
portions of the extended zone would help clarify the renormalization of the band 
structure in the CDW phase.

\section{Discussion.}%
The essence of our model lies in the three coupled effective 1D chains 
illustrated in \Fref{fig:structure}(b). Their weak hybridization entails a 
strongly anisotropic FS and, through hidden nesting and Coulomb 
interactions, determines a robust CDW instability 
\cite{whangbo1987band,whangbo1991hidden}.
The interaction parameters estimated in the RPA have magnitudes in the 
range ${\lesssim\,}1$\,eV expected for the octahedral MoO$_6$ network 
\cite{nuss2014effective} and are entirely consistent with the magnitudes 
needed to reproduce the experimental temperature dependence of the gap in 
$\Gamma K'$ (\Fref{fig:Gaps}).
The emergence of a band gap at this particular point in the BZ is not an 
obvious expectation \emph{a priori} (it is not nested by $\Qcdw$), and is 
another strong validating point.

That the experimental $\Tcdw$ and full $T$ dependence of the CDW gap are 
remarkably well described within a MF BCS-type theory (\Fref{fig:Gaps}) might 
seem unexpected at first given the reduced dimensionality. We attribute it to 
the combination of three factors: 
(i) Although the \emph{effective} 1D chains are a useful concept for the 
modeling, we saw that the actual system is quasi-2D given the nature of the 
electronic hybridization and interactions, which stabilizes the MF solution;
(ii) phase fluctuations, which tend to be the dominant suppressor of CDW order, 
are \emph{gapped} in a commensurate CDW \cite{gruner2000density};
(iii) the FS is fully gapped at $T{\,\le\,}\Tcdw$ except for the 
tiny pockets or arcs we find around $M'$ (\Fref{fig:BS-SpeFun-DOS}) and where ARPES 
reveals pseudogap-type behavior \cite{valbuena2006charge}.

Points (ii) and (iii) leave essentially no elementary nor collective excitations 
to destabilize the MF solution, indicating that the Ginzburg criterion may be well satisfied
over a large range of temperature below $\Tcdw$. Point (ii) is particularly important, in that it 
might explain not only the MF behavior of the bulk but also the strikingly 
different signatures of the gap attributed to the surface: A simple rescaling of 
the parameters cannot explain the secondary gap along $\Gamma K'$ seen in 
ARPES. 
This would seem to indicate that different microscopic details could be 
in play at the surface (this sometimes called ``extraordinary phase 
transition'' is common in other correlated systems \cite{brun2010surface, 
rosen2013surface}). A strongly-correlated state has been 
suggested \cite{mou2016discovery} but, not only is that in sharp contrast 
with the weak-coupling nature of this system in the bulk, it is incompatible 
with the experimental absence of any anomalous signatures in the normal state 
(including quasiparticle renormalization) other than the ``anomalous'' secondary 
gap. 
Another possibility, that we favor, is that fluctuations might be strongly 
enhanced at the surface. Figure~\ref{fig:Gaps}(a) shows that $\EgZero$ is very 
sensitive to the strength of interactions, as would be required to explain the 
higher stability of the CDW at the surface from reduced screening, but the 
experimental $T$ dependence is highly non-MF there. However, reduced screening 
combined with bolstered fluctuations can explain the downsizing in surface 
$\Tcdw$ in comparison with that (over)estimated in the MF based on the measured 
surface gap. The lock-in energy that drives commensurability is usually 
reinforced by interlayer CDW coupling, which will diminish for the surface 
slab. That can reduce the gap of the phase modes or even suppress it, since 
incommensurability can be favored under poor screening 
\cite{brown2005surface}, thus explaining the enhanced fluctuations. Current
experiments are not conclusive as to the (in)commensurability on the surface
but do reveal superlattice diffraction peaks much above the ``surface 
$\Tcdw$,'' albeit broadened and weak \cite{mou2016discovery}. This might point 
to phase fluctuations through discommensurations taking place at the surface, 
in line with the above picture.

\begin{acknowledgments}
We thank A.H. Castro Neto, C. Chen, and F. Hip\'olito for fruitful discussions.
The Singapore National Research Foundation supported this work with Grants No.
NRF-CRP6-2010-05 (L.S. and V.M.P.) and No. NRF-NRFF2013-03 (H.L.).
\end{acknowledgments}

\bibliography{CDW_in_KMoO} 



\section*{Supplementary material}

# — Supplementary Material —

# Charge Density Waves and the Hidden Nesting of Purple Bronze K$_{0.9}$Mo$_6$O$_{17}$

Lei Su,[1] Chuang-Han Hsu,[2,1] Hsin Lin,[2,1] and Vitor M. Pereira[2,1]

[1]*Centre for Advanced 2D Materials, National University of Singapore,*
*6 Science Drive 2, Singapore 117546*
[2]*Department of Physics, National University of Singapore,*
*2 Science Drive 3, Singapore 117542*

## COMPLETE 6-BAND TIGHT-BINDING MATRIX

The matrix $T_{\mu I, \nu J}(\boldsymbol{k})$ is given in the basis $\{\alpha A, \beta A, \gamma A, \alpha B, \beta B, \gamma B\}$ by

$$T(\boldsymbol{k}) = \begin{bmatrix} T_{AA}(\boldsymbol{k}) & T_{AB}(\boldsymbol{k}) \\ T_{AB}(\boldsymbol{k})^\dagger & T_{BB}(\boldsymbol{k}) \end{bmatrix} \tag{S1a}$$

with

$$T_{AA}(\boldsymbol{k}) = \begin{bmatrix} 2t_1 \cos(\boldsymbol{n}_1 \cdot \boldsymbol{k}) + \mu & t_2 & t_2 \\ t_2 & 2t_1 \cos(\boldsymbol{n}_2 \cdot \boldsymbol{k}) + \mu & t_2 \\ t_2 & t_2 & 2t_1 \cos(\boldsymbol{n}_3 \cdot \boldsymbol{k}) + \mu \end{bmatrix}, \tag{S1b}$$

$$T_{BB}(\boldsymbol{k}) = \begin{bmatrix} 2t_1 \cos(\boldsymbol{n}_1 \cdot \boldsymbol{k}) + \mu & t_2 e^{-i\boldsymbol{n}_3 \cdot \boldsymbol{k}} & t_2 e^{i\boldsymbol{n}_2 \cdot \boldsymbol{k}} \\ t_2 e^{i\boldsymbol{n}_3 \cdot \boldsymbol{k}} & 2t_1 \cos(\boldsymbol{n}_2 \cdot \boldsymbol{k}) + \mu & t_2 e^{-i\boldsymbol{n}_1 \cdot \boldsymbol{k}} \\ t_2 e^{-i\boldsymbol{n}_2 \cdot \boldsymbol{k}} & t_2 e^{i\boldsymbol{n}_1 \cdot \boldsymbol{k}} & 2t_1 \cos(\boldsymbol{n}_3 \cdot \boldsymbol{k}) + \mu \end{bmatrix}, \tag{S1c}$$

$$T_{AB}(\boldsymbol{k}) = \begin{bmatrix} t_0(e^{-i\boldsymbol{n}_2 \cdot \boldsymbol{k}} + e^{i\boldsymbol{n}_3 \cdot \boldsymbol{k}}) & t_3 e^{i\boldsymbol{n}_1 \cdot \boldsymbol{k}} & t_3 e^{-i\boldsymbol{n}_1 \cdot \boldsymbol{k}} \\ t_3 e^{-i\boldsymbol{n}_2 \cdot \boldsymbol{k}} & t_0(e^{i\boldsymbol{n}_1 \cdot \boldsymbol{k}} + e^{-i\boldsymbol{n}_3 \cdot \boldsymbol{k}}) & t_3 e^{i\boldsymbol{n}_2 \cdot \boldsymbol{k}} \\ t_3 e^{i\boldsymbol{n}_3 \cdot \boldsymbol{k}} & t_3 e^{-i\boldsymbol{n}_3 \cdot \boldsymbol{k}} & t_0(e^{-i\boldsymbol{n}_1 \cdot \boldsymbol{k}} + e^{i\boldsymbol{n}_2 \cdot \boldsymbol{k}}) \end{bmatrix}. \tag{S1d}$$

The vectors $\boldsymbol{n}_1 = (0, 2\sqrt{3}/3)$, $\boldsymbol{n}_2 = (1, -\sqrt{3}/3)$, and $\boldsymbol{n}_3 = (-1, -\sqrt{3}/3)$ are normalized to make $|\Gamma M| = \pi$. We have chosen to represent $T(\boldsymbol{k})$ in a gauge that reflects the $C_3$ symmetry more explicitly in the different hopping elements.

Fig. S1(a) is the *ab-initio* band structure. The full-potential first-principles calculations for the electronic structure of KMO were carried out with the Wien2k package. The generalized gradient approximation (GGA) [S1] was used for the exchange-correlation and a Monkhorst-Pack k-mesh of 10x10x10 was used for the self-consistent process. The convergence criterion of charge density and total energy were set below $10^{-6}$. We adopted the crystal structure with the unit cell and atomic positions obtained from the experiment[S2].

To obtain the TB parameters that best reproduce these *ab-initio* bands, we fit the bands of the former by varying its parameters until the sum of the squared residuals is minimized. We found that the signs and the relative magnitudes of the inter-orbital hoppings $t_2$ and $t_3$ are very important in determining the degeneracies of the three bottom bands at $\Gamma$ and K, in controlling the gap between the two degenerate bands and remaining one at $\Gamma$, and in fixing the intersection of the two lowest lying bands along $\Gamma$K. Given the crucial role

of the states near the Fermi surface in this system, the fit privileges the three partially filled bands by only looking at the energy window of 1 eV around $E_F$. Moreover, the best residual is not the deciding factor to select the best parameter set. Instead, the final TB parameters are those that provide the smallest residual while, *simultaneously* being able to reproduce the filling factor obtained *ab-initio*, the degeneracies at the high symmetry points and other band crossings, and the shape of the Fermi contours. We found the best such parameter set to be $\{t_0, t_1, t_2, t_3, \mu\} = \{454, -204, 136, 114, 659\}$ meV which generates the band structure represented in Fig. S1(b) [our calculations show, however, that once the intra-orbital hoppings are fixed, integrated quantities such as susceptibilities are not very sensitive to these minor features]. It is clear that the fit is not so good for the bands high above $E_F$ (by design, as per the above), but this is irrelevant given that the band renormalization taking place in the CDW phase is of the order of 50 meV.

The effect of the inter-chain hopping can be seen by comparison of this with Fig. S1(b), where we explicitly put $t_2 = t_3 = 0$ to electronically disconnect the three equivalent chains (this case is referred to as the "1D limit"). Panels (b) and (c) show that this is a quantitatively small effect, but qualitatively crucial in that, as seen in Fig. 2(a), it captures the Fermi surface measured in ARPES very well, and ultimately, determines the placement of the partial bandgaps and Fermi arcs in the CDW phase. ARPES suggests that inversion symmetry is partially broken even in the normal phase, [S3, S4] reducing the $C_6$ symmetry to $C_3$. This is possibly due to the slight distortion of the octahedra in the inner sub-layers of KMO [S5]. Our TB model, being strictly inversion symmetric, does not capture this, which is a small and not relevant feature for the description of the CDW instability.

We note also that, an alternative to the empirical TB approach that we develop here on the basis of symmetry and overlap of the $t_{2g}$ orbitals involved in the conduction bands, would be to determine maximally localized Wannier orbitals directly from the Bloch states obtained *ab-initio* [N. Marzari *et al.*, Rev. Mod. Phys. 84, 1419 (2012)], which is directly supported by Wien2k and its interface to Wannier90 [Mostofi *et al.*, Comput. Phys. Commun. 185, 2309 (2014)]. These would constitute best adapted basis for the tight-binding Hamiltonian, as far as agreement with the DFT bandstructure is concerned. However, given the good quality of our current parameterization for the three conduction bands, such calculation can only contribute a minor correction to our empirically formulated bands and, hence, yield only a marginal quantitative improvement to our current effective Hamiltonian. On the

3

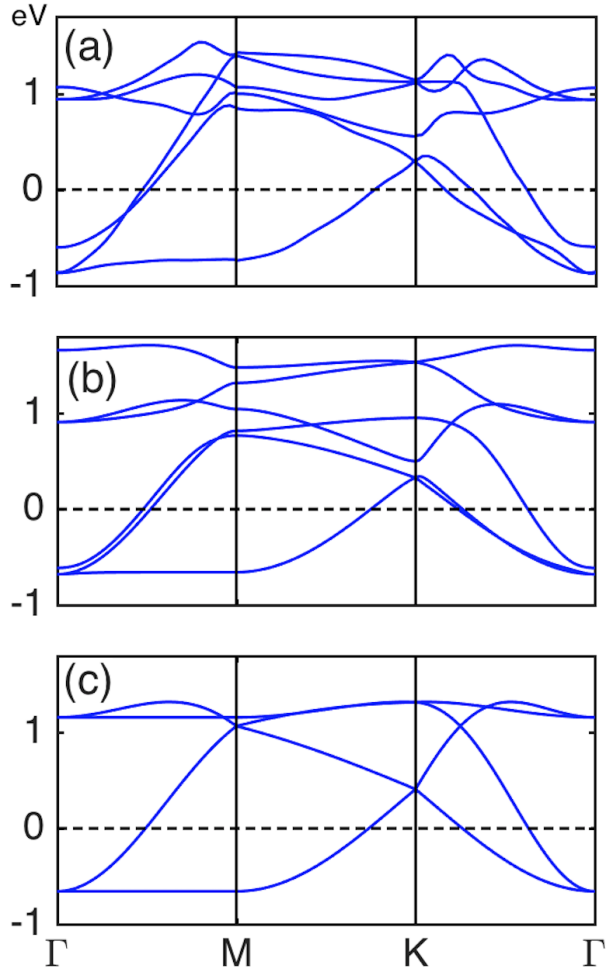

FIG. S1. (Color online) Band structures of KMO in the full BZ. (a) DFT calculated band structure; (b) band structure according to the TB model Eq. (S1a) with the parameters indicated in the text; (c) strictly 1D limit of the TB model ($t_2 = t_3 = 0$).

other hand, such an approach where the TB Hamiltonian appears as a by-product of the first principles bandstructure calculation (and requires it unavoidably) would obscure the intrinsic quasi one-dimensionality of this system which is transparent in our formulation. The more intuitive "construction" that arises in our description is thus not just a convenience, but is conceptually important for exposing the key underlying physics in a very direct way. This seems crucial for a successful model of electronic motion and interactions in this material and, for these reasons, we find it neither necessary nor advantageous to further refine the quantitative aspects of our TB in the context of this work.

**PARAMETRIZATION OF THE EXTENDED HUBBARD INTERACTIONS**

For a single orbital, e.g. $\alpha$, it is restricted within a zigzag chain and screening is expected to be incomplete. Since the band structure due to this orbital is well nested along the chain direction (apart from the hybridized portion), we assume that the intra-orbital interactions along this direction dominate. Without hybridization and other interactions, our model in this case may well reduce to an extended Hubbard model on a zigzag chain, which may be solved using standard techniques for 1D, such as, bosonization or renormalization group. Since neither spin density wave nor superconductivity is observed in KMO, we only have to study the charge instability. Moreover, since we know the instability is due to the nested Fermi surface at $\boldsymbol{Q}_\mu$ ($\mu \neq \alpha$ for $\alpha$-orbital), we are interested in the interaction at this particular wavevector. Consequently, the intra-sublattice interaction reduces to the ones usually considered for a 1D chain: $V_{\text{intra}} = U_{\text{onsite}} + V_{\text{nnn}} + \cdots$. As the on-site Hubbard repulsion acts only on electrons with different spins (and drives the spin density instability), it is likely that (the real part of) its Fourier transform is negative (at least for electrons with the same spin at $\boldsymbol{Q}_\delta$), i.e., $V_1(\boldsymbol{Q}_\delta) = -V_{\alpha I \alpha I}(\boldsymbol{Q}_\delta) = -(U_{\text{onsite}} - 2V_{\text{nnn}} + \cdots) > 0$. For the inter-sublattice component, however, (the real part of) the Fourier component may still be positive: $V_2(\boldsymbol{Q}_\delta) = -V_{\alpha I \alpha \bar{I}}(\boldsymbol{Q}_\delta) = -2V_{\text{nn}} \cos(\pi/3) < 0$, where we use $\bar{I}$ to denote a sublattice different from $I$ and the cosine term is due to the zigzag nature of the chain.

To go beyond these "1D" interactions between the same orbitals along each chain direction, we need to add both intra-orbital interactions perpendicular to each chain direction and inter-orbital interactions in all three directions. Focusing on one orbital, say $\alpha$, we set $V_1(\boldsymbol{Q}_\alpha) = \eta_1 V_1(\boldsymbol{Q}_\delta)$ for $\delta \neq \alpha$ (intra-orbital, same sublattice) and $V_2(\boldsymbol{Q}_\alpha) = \eta_2 V_2(\boldsymbol{Q}_\delta)$ for $\delta \neq \alpha$ (intra-orbital, different sublattice). The parameters $\eta_{1,2}$ model the anisotropy in the interaction for each orbital ($\alpha$ here). The other inter-orbital interactions are chosen to be $V_{\mu I \nu J}(\boldsymbol{Q}_\delta) = \eta_{\mu I \nu J, \delta} V_{\mu I \mu J}(\boldsymbol{Q}_\delta)$ for $\mu \neq \nu$, where $\eta_{\mu I \nu J, \delta}$ is a proportionality constant that expresses the *inter*-orbital interactions in terms of the *intra*-orbital terms, and allows one to keep a small number of total parameters. There is no way, in principle, to estimate these factors $\eta$ without detailed first-principles and exact diagonalization calculations; even with that information, it turns any subsequent calculation into a formidable task! Aiming at a good compromise between analytical simplicity and reproducibility of the key physics, we approximate them all [$\eta$ (inter), $\eta_1$ (intra), $\eta_2$ (intra)] by a single number $\eta$ in order to assess

the qualitative effect of these interactions. This leaves us with only three effective interaction parameters: $V_1$ capturing the inter-chain (quasi 1D) interactions, and $V_2$ capturing the interaction between adjacent chains. The factor $\eta$ determines the extent to which the full interacting Hamiltonian is more of 1D ($\eta=0$) or 2D character (large $\eta$). Other effects, such as differences that could arise in the relative importance of specific Coulomb terms for slabs located on the surface in comparison with those in the bulk are not considered at this point. Surface and bulk are distinguished within this model only by possible differences in the magnitude of the same interaction parameters. We do emphasize, however, that even though this estimation of interactions might appear crude, it does capture the essential physics qualitatively and quantitatively provided the intra-orbital interactions along the chain directions are well described due to the dominant role played by the Fermi surface nesting in this system. Our results showed in the main text supports this.

## SUSCEPTIBILITY AND INSTABILITY CRITERION

To assess the magnitude of the interactions capable of driving the system into the CDW phase, we first study the generalized Stoner criterion for this instability along the same lines used, for example, in multi-orbital iron-based superconductors [S6]. The generalized bare electronic susceptibility is given by

$$\chi^0_{\nu J,\mu I}(\boldsymbol{q},\omega) = \frac{1}{\mathcal{V}} \sum_{\boldsymbol{k},ll'} \frac{a_{l'}^{\mu I}(\boldsymbol{k})a_{l'}^{\nu J*}(\boldsymbol{k})a_{l}^{\nu J}(\boldsymbol{k}+\boldsymbol{q})a_{l}^{\mu I*}(\boldsymbol{k}+\boldsymbol{q})}{\omega + i0^+ + \xi_l(\boldsymbol{k}+\boldsymbol{q}) - \xi_{l'}(\boldsymbol{k})}(f_{l,\boldsymbol{k}+\boldsymbol{q}} - f_{l',\boldsymbol{k}}) \qquad \text{(S2)}$$

where $l, l'$ denote band indices after diagonalization of Eq. (S1a), $\xi_l(\boldsymbol{k})$ is the corresponding (band) energy eigenvalue, $f_{l,\boldsymbol{k}}$ is the Fermi-Dirac function, and $a_l^{\mu I}(\boldsymbol{k}) = \langle \mu I \boldsymbol{k} | l \boldsymbol{k} \rangle$ is the projection of the Bloch state onto the Bloch sum associated with the orbital $(\mu, I)$. In the RPA, the full and bare susceptibilities are related by

$$\chi_{\mu I,\nu J}(\boldsymbol{k}) = \chi^0_{\mu I,\nu J}(\boldsymbol{k}) - \chi^0_{\mu I,\eta K}(\boldsymbol{k}) V_{\eta K,\rho L}(\boldsymbol{k}) \chi_{\rho L,\nu J}(\boldsymbol{k}). \qquad \text{(S3)}$$

Since the $6 \times 6$ bare susceptibility matrix (S2) can be straightforwardly determined from the TB model, the generalized charge instability is given by the critical matrix of $V$ satisfying

$$\det\left\{ V(\boldsymbol{k}) + [\chi^0(\boldsymbol{k})]^{-1} \right\} = 0. \qquad \text{(S4)}$$

To begin, suppose we have only one zigzag chain, in which case we are reduced to a 2×2 problem with interaction and susceptibility matrices of the form

$$V(\boldsymbol{k}) = -\begin{pmatrix} V_1 & V_2 \\ V_2 & V_1 \end{pmatrix}, \qquad \chi^0(\boldsymbol{k}) = \begin{pmatrix} \chi_1^0 & \chi_2^0 \\ \chi_2^0 & \chi_1^0 \end{pmatrix}. \tag{S5}$$

It is then easy to see that the condition (S4) reduces explicitly to

$$(\chi_1^0 \pm \chi_2^0)(V_1 \mp V_2) = 1, \tag{S6}$$

from where we obtain $V_1 - V_2 \sim 800\,\text{meV}$ by putting $T = T_\text{c}$ in the expressions for $\chi_{1,2}^0$.

It is not difficult to generalize this approach to our actual system with three equivalent orbitals and a more complex interaction matrix $V$. The equivalence between the two sublattices dictates that the interaction must have the form of Eq. (S5) except that now $V_1$ and $V_2$ become 3×3 matrices, $\hat{V}_1$ and $\hat{V}_2$, and similar considerations apply to $\chi^0$ (the $\hat{\ }$ notation refers to the 3×3 matrices in the subspace spanned by the orbital indices):

$$V(\boldsymbol{k}) = -\begin{pmatrix} \hat{V}_1 & \hat{V}_2 \\ \hat{V}_2 & \hat{V}_1 \end{pmatrix}, \qquad \chi^0(\boldsymbol{k}) = \begin{pmatrix} \hat{\chi}_1^0 & \hat{\chi}_2^0 \\ \hat{\chi}_2^0 & \hat{\chi}_1^0 \end{pmatrix}. \tag{S7}$$

In our system, the off-diagonal block $\hat{\chi}_2^0$ of $\hat{\chi}^0$ is much smaller than the diagonal one. In this case, the generalized instability condition of Eq. (S4) becomes

$$\det\begin{pmatrix} \hat{V}_1 + [\hat{\chi}_1^0]^{-1} & \hat{V}_2 \\ \hat{V}_2 & \hat{V}_1 + [\hat{\chi}_1^0]^{-1} \end{pmatrix} = \det\{\hat{V}_1 + [\hat{\chi}_1^0]^{-1} + \hat{V}_2\}\det\{\hat{V}_1 + [\hat{\chi}_1^0]^{-1} - \hat{V}_2\} = 0, \tag{S8}$$

from where we identify the generalized Stoner criterion

$$\det\{\hat{V}_1 \pm \hat{V}_2 + [\hat{\chi}_1^0]^{-1}\} = 0. \tag{S9}$$

Knowledge of the interaction matrix $V$ permits the direct calculation of the transition temperature from this equation, assuming that the RPA describes the system well (which is often the case for a well-nested system [S7]). We follow a different strategy that consists in using Eq. (S9) to estimate the strength of the interaction parameters that is compatible with the experimental $T_\text{c}$ at the commensurate and experimentally robust $\boldsymbol{Q}_\mu$. For lack of detailed information about the interaction matrix, consider the limiting case where the interactions are isotropic regardless of orbital indices, sublattice indices and $\boldsymbol{Q}$ direction.

This corresponds to setting the parameter $\eta$ discussed above to $\eta = 1$, in which case the $3 \times 3$ blocks in the interaction matrix of Eq. (S7) become

$$\hat{V}_i = V_i \begin{pmatrix} 1 & 1 & 1 \\ 1 & 1 & 1 \\ 1 & 1 & 1 \end{pmatrix}, \qquad i \in \{1, 2\}. \tag{S10}$$

Eq. (S9) then reduces to another equation in $V_1 - V_2$, whose solution obtained with $\hat{\chi}_1^0(\boldsymbol{Q}_\alpha)$ computed from Eq. (S1) at $T = T_c$ yields $V_1 - V_2 \sim 280\,\text{meV}$. We take this result as a lower bound because of the coarseness of the above assumptions for the potential matrix (S10) that exaggerate the role of interactions and make the system more two-dimensional. Since the strictly 1D approximation discussed earlier using (S6) yielded the Stoner estimate $V_1 - V_2 \sim 800\,\text{meV}$, we expect the real system to have $280 \lesssim V_1 - V_2 \lesssim 800\,\text{meV}$. It follows that interactions are not too strong in comparison with the non-interacting bandwidths, and justifies *a posteriori* the analysis based on the RPA.

To illustrate the magnitude and momentum dependence of $\hat{\chi}_1^0$ at different temperatures, in Fig. 2(b) of the main text we plot its dominant matrix element, $\chi_d$. The dominant happens to always be along the diagonal, i.e., an intra-orbital and intra-sublattice susceptibility element, corresponding to one of the cases $\chi_{\nu I,\nu I}^0$ in eq. (S2). In that figure, we show the result for the two TB approximations discussed here, namely, the 1D approximation to the bandstructure where the 3 effective chains are decoupled (so, leaving 3 effectively decoupled 1D chains), and the full 2D model based on the TB fit to the *ab-initio* bandstructure. In each of these cases, the quantity plotted in Fig. 2(b) corresponds to one of the intra-sublattice ($\mu = \nu$) and intra-orbital ($I = J$) susceptibilities as follows:

$$\chi_d \equiv \chi_1^0 \qquad \text{[1D TB model, cf. (S5)]}, \tag{S11}$$

$$\chi_d \equiv \max\left\{\hat{\chi}_1^0\right\} \qquad \text{[2D TB model, cf. (S7)]}. \tag{S12}$$

Finally, we point out that the diagonal elements of $\hat{\chi}^0(\boldsymbol{q})$ (which are the elements of the type $\chi_{\nu I,\nu I}^0(\boldsymbol{q})$ in (S2) with $\nu \in \{\alpha, \beta, \gamma\}$) do not have the same $\boldsymbol{q}$ dependence along the directions defined by the CDW wavevectors $\boldsymbol{Q}_{\alpha,\beta,\gamma}$. This is obviously the case if one imagines the limit of uncoupled 1D chains: the chain labeled $\alpha$ in Fig. 1(b) (main text) runs vertically, parallel to $\boldsymbol{u}_y$; consequently, it has no dispersion along the direction of $\boldsymbol{Q}_\alpha$, but only along $\boldsymbol{Q}_\beta$ or $\boldsymbol{Q}_\gamma$. In this example, addition of the finite inter-chain hopping that is considered in

our realistic tight-binding introduces a small dispersion along $\boldsymbol{Q}_\alpha$, as expected. The point we wish to highlight with this is that, as a result of the asymmetry between inter-chain and intra-chain hoppings, for a given $\boldsymbol{q}$, the diagonal entries of $\hat{\chi}^0(\boldsymbol{q})$ will be different in general. The dominant one (highest in magnitude) is what we denote as $\chi_d$. In Fig. 2(b) (main text) we are interested in the momentum dependence along ΓM because the CDW instability occurs at one of the M points. We thus choose the direction parallel to one specific $\boldsymbol{Q}_\mu$ to represent $\chi_d$. Obviously, the choice of a particular $\boldsymbol{Q}_\mu$ is arbitrary, as dictated be the $C_3$ symmetry of the system, and a different choice will only select a correspondingly different entry of $\hat{\chi}^0(\boldsymbol{q})$ as the dominant one, but the momentum dependence of the new $\chi_d(\boldsymbol{q})$ along the new $\boldsymbol{Q}_\mu$ remains exactly the same.

**HARTREE-FOCK ANALYSIS**

Having established the weak coupling nature of this problem and the relatively high transition temperature, we hope to capture the full temperature dependence of the CDW order parameter and the band structure renormalization with a mean-field (MF) analysis of the interacting Hamiltonian defined by Eqs. (S1) and (2). Its Hatree-Fock decoupling yields

$$H = H_0 - \frac{1}{2\mathcal{V}} \sum_{\boldsymbol{kQ}} \sum_{\mu\nu I} [V_{1,\mu\nu}(\boldsymbol{Q})\Delta_{\mu I}(-\boldsymbol{Q})c^\dagger_{\nu I\boldsymbol{k}+\boldsymbol{Q}}c_{\nu I\boldsymbol{k}} + V_{1,\mu\nu}(\boldsymbol{Q})\Delta_{\nu I}(\boldsymbol{Q})c^\dagger_{\mu I\boldsymbol{k}-\boldsymbol{Q}}c_{\mu I\boldsymbol{k}}]$$
$$- \frac{1}{2\mathcal{V}} \sum_{\boldsymbol{kQ}} \sum_{\mu\nu I} [V_{2,\mu\nu}(\boldsymbol{Q})\Delta_{\mu\bar{I}}(-\boldsymbol{Q})c^\dagger_{\nu I\boldsymbol{k}+\boldsymbol{Q}}c_{\nu I\boldsymbol{k}} + V_{2,\mu\nu}(\boldsymbol{Q})\Delta_{\nu I}(\boldsymbol{Q})c^\dagger_{\mu\bar{I}\boldsymbol{k}-\boldsymbol{Q}}c_{\mu\bar{I}\boldsymbol{k}}]$$
$$+ \frac{1}{2\mathcal{V}} \sum_{\boldsymbol{Q}} \sum_{\mu\nu I} V_{1,\mu\nu}(\boldsymbol{Q})\Delta_{\mu I}(-\boldsymbol{Q})\Delta_{\nu I}(\boldsymbol{Q}) + \frac{1}{2\mathcal{V}} \sum_{\boldsymbol{Q}} \sum_{\mu\nu I} V_{2,\mu\nu}(\boldsymbol{Q})\Delta_{\mu\bar{I}}(-\boldsymbol{Q})\Delta_{\nu I}(\boldsymbol{Q}), \quad \text{(S13)}$$

where $\bar{I} \neq I$, $\Delta_{\mu I}(\boldsymbol{Q}) \equiv \sum_{\boldsymbol{k}} \langle c^\dagger_{\mu I\boldsymbol{k}+\boldsymbol{Q}}c_{\mu I\boldsymbol{k}}\rangle$, $\boldsymbol{Q} \in \{\pm\boldsymbol{Q}_\alpha, \pm\boldsymbol{Q}_\beta, \pm\boldsymbol{Q}_\gamma\}$. Since $2\boldsymbol{Q}$ is a reciprocal lattice vector and $\Delta_{\mu I}(\boldsymbol{Q}) = \Delta^*_{\mu I}(-\boldsymbol{Q})$, we can choose $\Delta_{\mu I}(\boldsymbol{Q})$ to be real and rewrite

$$H = H_0 - \frac{1}{\mathcal{V}} \sum_{\boldsymbol{kQ}} \sum_{\mu\nu I} \left[ [V_{1,\mu\nu}(\boldsymbol{Q})\Delta_{\mu I}(\boldsymbol{Q}) + V_{2,\mu\nu}(\boldsymbol{Q})\Delta_{\mu\bar{I}}(\boldsymbol{Q})] c^\dagger_{\nu I\boldsymbol{k}+\boldsymbol{Q}}c_{\nu I\boldsymbol{k}} \right]$$
$$+ \frac{1}{2\mathcal{V}} \sum_{\boldsymbol{Q}} \sum_{\mu\nu I} [V_{1,\mu\nu}(\boldsymbol{Q})\Delta_{\mu I}(\boldsymbol{Q}) + V_{2,\mu\nu}(\boldsymbol{Q})\Delta_{\mu\bar{I}}(\boldsymbol{Q})]\Delta_{\nu I}(\boldsymbol{Q}), \quad \text{(S14)}$$

9

where $V_{i,\mu\nu}(\boldsymbol{Q})$ has the same magnitude for all $\boldsymbol{Q}$ The first line of the Hamiltonian above can be written in an explicit form as

$$\begin{pmatrix} T_{\boldsymbol{k}} & W_\alpha & W_\beta & W_\gamma \\ W_\alpha & T_{\boldsymbol{k}+\boldsymbol{Q}_\alpha} & W_\gamma & W_\beta \\ W_\beta & W_\gamma & T_{\boldsymbol{k}+\boldsymbol{Q}_\beta} & W_\alpha \\ W_\gamma & W_\beta & W_\alpha & T_{\boldsymbol{k}+\boldsymbol{Q}_\gamma} \end{pmatrix} \tag{S15}$$

where each entry is a 6×6 block: $T_{\boldsymbol{k}}$ is the TB matrix (S1) with $\boldsymbol{k}$ in the reduced first BZ, and

$$W_\delta \equiv -\frac{2}{\mathcal{V}} \sum_{\mu\nu I} V_{1,\mu\nu}(\boldsymbol{Q}_\delta)\Delta_{\mu I}(\boldsymbol{Q}_\delta) + V_{2,\mu\nu}(\boldsymbol{Q}_\delta)\Delta_{\mu \bar{I}}(\boldsymbol{Q}_\delta). \tag{S16}$$

At zero temperature, minimization of the ground state energy of the Hamiltonian (S14) with respect to the order parameter $\Delta_{\mu I}(\boldsymbol{Q})$ leads to the following electronic density contributed by orbital $\mu$:

$$\rho_\mu(\boldsymbol{R}_i) = \rho_\mu(0) + \frac{1}{\mathcal{V}} \sum_{\boldsymbol{Q}} [e^{i\boldsymbol{R}_i \cdot \boldsymbol{Q}}(e^{i\boldsymbol{\delta}_A \cdot \boldsymbol{Q}}\Delta_{\mu A}(\boldsymbol{Q}) + e^{i\boldsymbol{\delta}_B \cdot \boldsymbol{Q}}\Delta_{\mu B}(\boldsymbol{Q}))], \tag{S17}$$

and

$$\delta\rho_\mu(\boldsymbol{R}_i) = \frac{1}{2\mathcal{V}} \sum_{\boldsymbol{Q}} [\cos(\boldsymbol{R}_i \cdot \boldsymbol{Q} + \boldsymbol{\delta}_A \cdot \boldsymbol{Q})\Delta_{\mu A}(\boldsymbol{Q}) + \cos(\boldsymbol{R}_i \cdot \boldsymbol{Q} + \boldsymbol{\delta}_B \cdot \boldsymbol{Q})\Delta_{\mu B}(\boldsymbol{Q})]. \tag{S18}$$

At finite temperatures, the stable solution is obtained by minimizing the electronic free energy given by

$$F = -T \sum_{\boldsymbol{k}} \sum_{l} \ln(1 + e^{-\xi_l(\boldsymbol{k})/T}) + G(\Delta), \tag{S19}$$

where $T$ is the temperature (in units of energy), $G(\Delta)$ is the second line in Eq. (S14), and $\xi_l(\boldsymbol{k})$ are the renormalized band dispersions. Once the temperature, TB parameters and interactions are specified, the finite-temperature MF solution is obtained by numerical minimization of Eq. (S19). For interactions in the 1D limit, we find that the order parameters display the following general features: (i) they always have opposite sign for the same orbital but different sublattice, $\Delta_{\mu I}(\boldsymbol{Q}_\nu) = -\Delta_{\mu \bar{I}}(\boldsymbol{Q}_\nu)$; (ii) for a given chain, they vanish for $\boldsymbol{Q}$ perpendicular to that chain's direction, e.g., $\Delta_{\alpha I}(\boldsymbol{Q}_\alpha) = 0$; (iii) for a given chain, they are equal for $\boldsymbol{Q}$ along the direction of the other two, e.g., $\Delta \equiv \Delta_{\alpha I}(\boldsymbol{Q}_\beta) = \Delta_{\alpha I}(\boldsymbol{Q}_\gamma) \neq 0$.

**PERTINENT DIRECTIONS FOR THEORETICAL REFINEMENTS**

Even though the behavior of the bulk has been shown to be well described by a mean-filed type treatment of the Coulomb interactions, we discuss two natural extensions of our theoretical analysis that could be relevant to illuminate the underlying origin of this agreement, as well as the puzzling experimental observations in the surface layer.

**Landau Theory**

While MF theories of CDW usually fail near $T_c$ [S8], the good agreement reported here is likely a result of the true 2D nature of the electronic system in KMO that, albeit weak, combined with the fully 3D dielectric environment contributes to stabilizing the order parameter against fluctuations. An additional or alternative contributor might be the strictly commensurate nature of the CDW which generically gaps phase fluctuations. To assess this in more detail, one may attempt to derive the phenomenological theory from the microscopic theory, but it should be reminded that the effective Hamiltonian is written in terms of Wannier orbitals. Since the CDW phase is commensurate at half of the reciprocal lattices, the $\Delta(\boldsymbol{Q}_\mu)$ is forced to be real. The quadratic term of the Landau theory is given by the form: $\mathcal{S}^{(2)} = 1/[2T\mathcal{V}] \sum_{\boldsymbol{Q}_\mu} \Delta^{\mathrm{T}}(\boldsymbol{Q}_\mu)[V^{-1}(\boldsymbol{Q}_\mu) - \chi(\boldsymbol{Q}_\mu)]\Delta(\boldsymbol{Q}_\mu)$ where $\Delta$ consists the six order parameters for each orbital and sublattice. The quartic terms of the long wave fluctuations around $\boldsymbol{Q}_\mu$ can be written down using loop expansions. Notice, however, that at least to the MF level, the phase of the charge density, which depends on $\langle c^\dagger_{\mu I \boldsymbol{k}+\boldsymbol{Q}} c_{\mu I \boldsymbol{k}} \rangle$ for each $\boldsymbol{k}$ (rather than the sum), does not play an explicit role, unlike the Landau theory derived using the Fermi liquid picture [S8]. This approach can also be compared with McMillan's theory for conventional CDW [S9] where $\delta\rho$ is a function of $\boldsymbol{r}$ instead of $\boldsymbol{R}_i$ (Eq. (S18)).

**Refinements in the treatment of interactions**

The very large and complex unit cell of KMO make a more detailed treatment of the electron-electron interactions a rather challenging problem, at least in terms of effective models. Given this intrinsic complexity, should refinement of the microscopic details be necessary (for example, in the wake of new experiments and new details of the behavior at the surface), it seems desirable that they be tackled directly with resort to reliable numerical

techniques. Full self-consistent calculations that combine DFT and dynamical mean-field theory (DMFT) might be pursued in the general frameworks of ABINIT [S10] or TRIQS [S11, S12]. Variations of the DMFT concept such as the Cluster Dynamical Mean-Field Theory or the Variational Cluster Approximation might provide more precise details of the degree of correlations in the system [S13].

Whether the 2D nature of interactions is indispensable (as is implicitly the case for the system to select the measured $\boldsymbol{Q}_\mathrm{cdw}$ which does not nest any of the underlying effective 1D Fermi contours, and is explicitly the case within our proposed model to obtain the best experimental agreement, as discussed in relation to Fig. 3, for example) could be addressed with methods that are either fundamentally applicable to 1D systems, or that allow a controlled, perturbative departure from 1D. Approaches based on Tomonaga-Luttinger liquid theory or Density Matrix Renormalization Group (DMRG) [S14] are the natural directions in this regard, and they have been pursued in the context of the related Li bronze $\mathrm{Li}_{0.9}\mathrm{Mo}_6\mathrm{O}_17$ [S15, S16]. However, it should be underlined that the Li bronze is electronically quite different, in particular, having electronic structure and interactions with a strongly one-dimensional character [S17]. In the case of KMO, the underlying one-dimensionality as seen from the viewpoint of the hidden nesting might be misleading since the 2D nature of the interactions seems to be important, and the 1D limit might not be as good a starting point here as in the case of Li.

———


\end{document}